\documentclass{kapproc} 

\RequirePackage{graphicx}%
\RequirePackage{epsf}%
\input{psfig}

\upperandlowercase
\let\footnote\savefootnote
\let\footnotetext\savefootnotetext 
 
\kluwerbib 

\def\BP{Ballesteros-Paredes}
\def\gtsima{$\; \buildrel > \over \sim \;$}    
\def\gtrsim{\lower.5ex\hbox{\gtsima}}           
\def\ltsima{$\; \buildrel < \over \sim \;$}    
\def\lesssim{\lower.5ex\hbox{\ltsima}}           
\def\Lj{L_{\rm J}}
\def\Ljc{L_{\rm J,c}}
\def\Ljg{L_{\rm J,g}}
\def\ls{\lambda_{\rm s}}
\def\tfc{\tau_{\rm f,c}}
\def\tad{\tau_{\rm AD}}
\def\tfg{\tau_{\rm f,g}}
\def\VS{V\'azquez-Semadeni}

\begin{document}


\articletitle{Turbulent Control of the Star \\
Formation Efficiency}


\chaptitlerunninghead{Turbulent Control of Star Formation Efficiency}



 \author{Enrique V\'azquez-Semadeni}
 \affil{Centro de Radioastronom\'ia y Astrosf\'isica, UNAM,
Morelia, M\'exico.}
 \email{e.vazquez@astrosmo.unam.mx}





 \begin{abstract}
Supersonic turbulence plays a dual role in molecular
clouds: On one hand, it contributes to the global support of the clouds,
while on the other it promotes the
formation of small-scale density fluctuations, identifiable with clumps
and cores. Within these, the local Jeans length $\Ljc$ is reduced, and collapse
ensues if $\Ljc$ becomes smaller than the clump size
and the magnetic support is insufficient (i.e., the core is ``magnetically
supercritical''); otherwise, the clumps do not collapse 
and are expected to re-expand and disperse on a few free-fall
times. This case may correspond to a fraction of the observed starless
cores. The star formation 
efficiency (SFE, the fraction of the cloud's mass that ends up in
collapsed objects) is smaller than unity because the mass contained
in collapsing clumps is smaller than the total cloud
mass. However, in non-magnetic numerical simulations with realistic
Mach numbers and turbulence driving scales, the SFE is still
larger than observational estimates. The presence of a magnetic
field, even if magnetically supercritical, appears to further reduce the
SFE, but by reducing the probability of core formation rather than by
delaying the collapse of individual cores, as was formerly thought. Precise
quantification of these effects as a function of global cloud parameters
is still needed. 

 \end{abstract}

\section{Introduction}

The observed efficiency of star formation (SFE, the
fraction of a cloud's mass deposited in stars during its lifetime) is
low, on the order of a few percent (e.g., \cite{Evans91}). For over
two decades, the accepted explanation (Mouschovias 1976a,b;
\cite{SAL87}) to this low observed SFE was that low-mass stars form in
so-called ``magnetically subcritical'' molecular clouds, which, under ideal
MHD conditions (perfect flux freezing), would be absolutely
supported by magnetic forces against their own self-gravity, regardless
of the external pressure. 
%
%
In practice, however, in dense clumps within the clouds, the 
ionization fraction drops to sufficiently low values that 
the process known as ``ambipolar diffusion'' (AD; \cite{Mes_Spit56})
allows quasi-static contraction of the clumps into denser structures
(``cores''), and ultimately collapse.
%
The low SFE then arises from the fact that only the material in the
densest regions could proceed to gravitational collapse, and on the
AD time scale, which is in general much larger than the
free-fall time scale. High-mass stars, on the other hand, were proposed
to form from 
either supercritical clouds assembled by agglomeration of smaller clouds
into large complexes (\cite{SAL87}), or by super-Alfv\'enic shock
compression of sub- or nearly critical clouds (\cite{Mousch91}).
In this scenario, which we refer to as the ``standard model'' of star
formation, {\it gravitational} fragmentation along flux
tubes containing many Jeans masses (e.g., Shu et al.\ 1987;
\cite{Mousch91}), was considered to be the mechanism responsible for
clump formation. 

However, molecular clouds are known to be supersonically turbulent
(e.g., \cite{ZE74}; \cite{Larson81}; \cite{Blitz93}), 
and this is bound to produce large density fluctuations, even
if the turbulence is sub-Alfv\'enic.\footnote{In fact, numerical
simulations suggest that strongly magnetized cases develop larger
density contrasts than weakly magnetized ones (e.g., \cite{PVP95};
\cite{OGS99}; \cite{BP_ML02}).} In this case, the clumps and cores
within molecular clouds, as well as the clouds themselves, are
likely to be themselves the turbulent density fluctuations within the
larger-scale turbulence of their embedding medium (\cite{vonWeiz51};
\cite{Sasao73}; \cite{Elm93}; \cite{BVS99}), being transient,
time-dependent, out-of-equilibrium objects in which the kinetic
compressive energy of the large-scale turbulent motions is being
transformed into the internal, gravitational and perhaps
smaller-scale turbulent kinetic energies of the density enhancements. 
The typical formation time scales of the density fluctuations should be
of the order of the rms turbulent crossing time across them.

If clumps and cores within molecular clouds are indeed formed through this
rapid, dynamic process, such an origin and out-of-equilibrium nature 
appear difficult to reconcile with the quasi-magnetostatic nature of the
AD contraction proposed to occur in the standard
model. Moreover, a number of additional problems with the standard model
have been identified (see the review by \cite{ML_Kl04}), among which a
particularly important one is that molecular clouds are generally
observed to be magnetically supercritical or nearly critical (e.g.,
Crutcher 1999, 2003; \cite{Bourke_etal01}), in agreement with
expectations from the cloud formation mechanism (\cite{HBB01}).

In this paper we review how the SFE can be maintained at low levels
within the context of what has become known as the ``turbulent model''
of molecular cloud formation, without having to necessarily resort to
quasi-static, AD-mediated slow contraction.

\section{Turbulent control of gravitational collapse}

In this section we consider the role of turbulence neglecting the
magnetic field.

In most early treatments of self-gravitating clouds, turbulence had
been considered only as a source of support (e.g., \cite{Chandra51};
\cite{Bonaz87}; \cite{Liz_Shu89}). However, one of the main 
features of turbulence is that it is a multi-scale process, with most of
its energy at large scales. Thus, it
is expected to have a {\it dual} role in the dynamics of molecular
clouds (\cite{VP99}): over all scales on which turbulence is supersonic, 
it is the dominant form of support, while simultaneously it induces the
formation of small-scale density peaks (``clumps''). If the latter are still
supersonic inside, further, smaller-scale peak formation is expected in
a hierarchical 
manner (\cite{VS94}; \cite{PV98}), until small enough scales are
reached that the typical velocity differences across them are subsonic, at
which point turbulent energy ceases to be dominant over thermal energy
for support, and also further turbulent subfragmentation cannot occur
(\cite{Padoan95}; \cite{VBK03}). These can be identified (\cite{KBVD04})
with ``quiescent'' (\cite{Myers83}), ``coherent'' (\cite{Goodman_etal98})
cores. We refer to the scale at which the typical turbulent
velocity difference equals the sound speed as the ``sonic scale'',
denoted $\ls$. It
depends on both the slope and the intercept of the turbulent energy
spectrum. 

For a molecular cloud of mean density $n \sim 10^3$
cm$^{-3}$ and temperature 10 K, the thermal Jeans length is $\Lj 
\sim 0.7$ pc, and so sub-parsec clumps will generally be smaller than the
cloud's {\it global} Jeans length $\Ljg$. However, in the clumps, the {\it
local} Jeans 
length $\Ljc$ is reduced, and in some cases it may become smaller
than the clump's size, at which point the clump can proceed to collapse.  
If the clump is internally subsonic, then $\Ljc$ is given by the
thermal Jeans length; otherwise, $\Ljc$ should include the turbulent
support (\cite{Chandra51}). Moreover, in the latter case, the clump can 
still fragment due to the turbulence, with the fragments collapsing
earlier than the whole clump (because they have shorter free-fall
times), probably producing a bound cluster. On the other hand, if
$\Ljc$ never becomes smaller than clump's size during the compression,
then the clump is expected to re-expand 
after the turbulent compression ends, on times a few times larger
than the free-fall time (\cite{VKSB05}). This is because in the
absence of a magnetic field, {\it stable} equilibria of self-gravitating
isothermal spheres require the presence of an external, 
confining pressure. In
the case of clumps formed as turbulent fluctuations, the
external pressure includes the fluctuating turbulent ram pressure, and
is therefore time variable, being at a maximum when the clump is being
formed, and later returning to the mean value of the ambient thermal
pressure.  

The formation of collapsing objects is a highly nonlinear and time
dependent process,
which is most easily investigated numerically. Early studies in two
dimensions suggested that gravitational collapse could be almost completely
suppressed by turbulence driving if the driving was applied at scales
smaller than the global Jeans length (\cite{LPP90}). This was 
later supported by the 3D studies of \cite{KHM00}, who investigated
the evolution of the collapsed mass fraction as a function of the rms
Mach number and the driving scale of the turbulence in numerical
simulations of isothermal, non-magnetic, self-gravitating driven
turbulence. However, for driving scales larger than $\Ljg$,
Klessen et al.\ (2000) still found collapsed fractions well below
unity, showing that the SFE is reduced 
even if the driving scale is larger than $\Ljg$. This is
important because it is likely that the turbulence in molecular clouds
is driven from large scales (\cite{Oss_ML02}), the clouds actually
being part of the general turbulent cascade in the ISM (\VS\ et al.\
2003).  In this case, the driving scale is not a free parameter, and
the ability to reduce the SFE even with large-scale driving is essential.

\VS\ et al.\ (2003) later showed that, for the simulations they
considered, all 
of which had the same number of Jeans lengths in the box ($J$, equal
to 4 there), the SFE correlates better with the
sonic scale than with either the rms Mach number or the driving scale,
substantiating the relevance of the sonic scale. The correlation was
empirically fit to a function of the form SFE $\propto
\exp{(-\lambda_0/\ls)}$, with $\lambda_0 \sim 0.11$ pc in the
simulations studied. If the driving scale is
kept constant (say, at its largest possible value), then the
dependence of the SFE on the sonic scale translates directly into a
dependence on the Mach number. Indeed, the data of Klessen et al.\
(2001) and of \VS\ et al.\ (2003) show that the SFE in simulations driven 
at a fixed scale is systematically reduced as the Mach number is
increased. For example, in simulations driven at $2 \Ljg$,
efficiencies $\sim$ 40\% were observed for rms Mach numbers $\sim
10$ (\VS\ et al.\ 2003). A theory explaining this functional
dependence is lacking. Moreover, the experiments
so far remain incomplete, since they have not tested the dependence of
the SFE on the Jeans number of the flow $J$.

In summary, in the non-magnetic case, the numerical experiments show
that the SFE can be reduced by turbulence alone, without the need for
magnetic fields. However, for realistic rms Mach numbers, the efficiencies 
observed are still larger than observed if one admits that clouds are
likely to be driven at large scales.

\section{The role of the magnetic field}

The magnetic field may provide the necessary further reduction of the
SFE to reach the observed levels, even in supercritical clouds. Early
numerical studies showed that global collapse in a magnetic simulation
can only be 
completely suppressed if the box is magnetically subcritical and AD is
neglected (Ostriker et al.\ 1999; \cite{HMK01}). Supercritical boxes
readily collapse, although on time scales up to twice the global free-fall 
time $\tfg$. Heitsch et al.\ (2001) and Li et al.\ (2004) additionally
have shown that MHD waves within supercritical clumps are apparently
insufficient to prevent their collapse. Heitsch et al.\ (2001) also
investigated the collapsed mass fraction as a function of magnetization in
supercritical boxes, but were not able to find any clear trends, because
the effect of the magnetic field was obscured by stochastic variations
between different realizations of flows with the same global parameters.

Recently, \cite{LN04} have considered the same issue in two-dimensional,
decaying (rather than driven) simulations, in both sub- and
supercritical regimes, including a prescription for
AD. They found that stronger magnetic fields delay the initial
formation of collapsed objects, although all their simulations
at a fixed Mach number achieved comparable final collapsed 
mass fractions at long times. They also concluded that higher levels of
initial turbulence speed up the collapse in subcritical clouds by
producing high-density clumps in which the AD time scale is short, and
thus avoiding the problem that AD by itself gives collapse times that
are too long compared to observational evidence (e.g., \cite{JMA99};
\cite{LM99}; \cite{Hartmann03}). 

The above studies have focused on the global collapsed mass fraction in
simulations, but further insight can be obtained by focusing on the
evolution of individual clumps. \VS\ et al. (2005) have
investigated the evolution of individual clumps in three-dimensional,
driven MHD 
simulations neglecting AD. These showed that the typical times for clumps to 
go from mean densities $\sim 10^4$ cm$^{-3}$ (the level of the densest
fluctuations produced by the turbulence in their Mach-10 simulations)
to full collapse differ by less than a factor 
of 2 between supercritical and non-magnetic simulations, being $\lesssim
2 \tfc \sim 1$ Myr in the former, and $\sim 1 \tfc \sim 0.5$ Myr in the
latter, where  
$\tfc\equiv \Ljc/c$ is the {\it local} free-fall time in the clumps, and $c$
is the isothermal sound speed. Furthermore, these authors showed that
in subcritical simulations without AD, in which collapse cannot occur, 
the clumps only reached mean densities $\sim 10$--$20 \times 10^4$
cm$^{-3}$, to then rapidly become dispersed again in times $\sim 1$
Myr. An estimate of the AD time scale $\tad$ in one such clump taking
into account its closeness to the critical mass-to-flux ratio
(\cite{CB01}) gave $\tad \gtrsim 1.3$ Myr, suggesting that in the presence
of AD the clump might possibly increase its mass-to-flux ratio and
proceed to collapse by the effect of AD, although on time scales not
significantly longer than the dynamical ones observed in the
supercritical and non-magnetic simulations. If AD acts on significantly
longer time scales, then it cannot bind the clumps before they are dispersed
by the turbulence.

\VS\ et al. (2005) also noticed that the appearance
of the first collapsing cores in the supercritical simulations was
delayed with respect with the non-magnetic simulation, and that
fewer cores formed in the magnetic cases than in the non-magnetic
one. These findings are consistent with previous results that the presence 
of the magnetic field delays the collapse (Ostriker et al.\ 1999;
Heitsch et al.\ 2001), but suggests that the delay at the global scale
occurs by reducing the probability of forming collapsing cores, rather
than by delaying the collapse of individual clumps. This may be the
consequence of the magnetic field reducing the effective dimensionality
of turbulent compressions, which become nearly one-dimensional in the
limit of very strong fields, in which case the compressions cannot
produce collapsing objects (e.g., \cite{SAL87}; \cite{VPP96}).

\section{Conclusions}

The results summarized here show that the SFE in supersonically
turbulent molecular clouds is naturally reduced because the turbulence
opposes global cloud collapse while inducing the formation of
local density peaks that contain small fractions of the total mass, and
which may collapse if they become locally gravitationally
unstable. However, not all density peaks (``clumps'') manage to do so,
and a number of them are expected to instead re-expand and merge with
their environment. This mechanism operates even in the absence of
magnetic fields, although for realistic parameters of the turbulence,
the efficiencies in numerical simulations are higher than
observed. Including the magnetic field further reduces the efficiency of
collapse, even in supercritical cases, but apparently not by delaying
the formation and collapse of individual clumps, which occurs on
comparable time scales in both the magnetic and non-magnetic cases, but
by reducing the probability of collapsing-core formation by the
turbulence. Further work is now needed to quantify the SFE and the
fraction of collapsing versus non-collapsing peaks as a
function of the global parameters, and to eventually produce a collective
theory that describes the process in a statistical fashion.

\begin{acknowledgments}
The author has benefitted from a critical reading of the manuscript by
Javier Ballesteros-Paredes, and from financial support from CONACYT
through grant 36571-E.
\end{acknowledgments}

\end{document}